\documentclass{revtex4}
\usepackage{amssymb}

\usepackage{amsmath}
\usepackage{graphicx}
\usepackage{amsmath}

\usepackage[colorlinks]{hyperref}

\newtheorem{theorem}{Theorem}
\newtheorem{acknowledgement}[theorem]{Acknowledgement}

\begin{document}

\title{On the equation of state of a flat FRW model filled with a bulk viscous
fluid.}
\author{J.A. Belinch\'{o}n}
\email{abelcal@ciccp.es}
\affiliation{Dpt. of Physics ETS
Architecture. UPM Madrid. Av. Juan de Herrera N-4 28040 Espa\~na.}
\date{\today}

\begin{abstract}
In this paper, we study the equation of state admissible for a flat FRW
models filled with a bulk viscous fluid by using the Lie group method. It is
found that the model admits scaling symmetries iff the bulk viscous
parameter $\gamma =1/2$. In this case, it is found that the main quantities
follow a power law solution and in particular the bulk viscous pressure $\Pi
$ has the same order of magnitude as the energy density $\rho ,$ in such a
way, that it is possible to formulate the equation of state $\Pi =\varkappa
\rho ,$ where $\varkappa \in \mathbb{R}^{-}$ (i.e. is a negative numerical
constant)$.$ If we assume such relationship we find again that the model is
scale invariant iff $\gamma =1/2.$ We conclude that the model accepts a
scaling symmetry iff $\gamma =1/2$ and that for this value of the viscous
parameter, $\Pi =\varkappa \rho ,$ but the hypothesis $\Pi =\varkappa \rho $
does not imply $\gamma =1/2,$ and that the model is scale invariant.
\end{abstract}

\maketitle

\section{Introduction}

As it is known in General Relativity (GR) there is a dualism between
spacetime and matter. While the structure of the spacetime is governed by
the field equations the physical properties of matter are introduced through
the energy-momentum tensor attending to diverse physical considerations.
Some of these considerations come from fields of the physic where gravity
does not play any role and are assumed in GR. As the Einstein equations
together Bianchi identities form an undetermined system of equations, it is
necessary to introduce equation of state, some of them ad hoc, in such a way
that the resulting system of equations may be integrated. Such system
satisfy certain symmetries that form a group, the group of symmetries of the
equations.

Collins (\cite{Collins}) and later on M. Szydlowsky (\cite{Polacos}, we
follow closely this work) have used the inverse way in order to determine
the admissible equations of state for a system of equations under the
restriction that this system admits a determined group of symmetries i.e.
one could a priori assume a symmetry group of the Einstein equations and out
of it deduce the condition of integrability which has the form of the
equation of state. In this way the group of symmetries of the Einstein
equations correctly select physical meaningful equation of state.

Since the bulk viscous theory is constructed assuming phenomenological (ad
hoc) equations of state and therefore the equations depend on certain
undetermined numerical constants, we are interested in determining the exact
form of such equations of state imposing the condition that the field
equations admit a concrete symmetry. Therefore, in this paper, we show that
the field equations of a cosmological flat FRW model filled with a bulk
viscous fluid together with a suitable equations of state admits a certain
Lie group of symmetries or, vice versa, the invariance of the field
equations with respect to a given symmetry group singles out the
corresponding equation of state.

The paper is divided as follows. In section 2 we outline the general field
equations of our model i.e. a flat FRW bulk viscous fluid and without the
cosmological constant. In section 3 we deduce the second order ode that
governs the model. This differential equation has been deduced without any
assumption. One we have outlined the basic ode, using the Lie group
technique we study this equation finding that only admits one symmetry but
if we impose that the model admits the scaling symmetry then we find that
this is only possible if the viscous parameter $\gamma =1/2$ (where the
viscosity $\xi $ has been introduced into the field equations through the ad
hoc law $\xi =k_{\gamma }\rho ^{\gamma },$ and in particular, we are
interested in determining the possible value(s) of the parameter $\gamma $).
Once we have established that the field equations are scale-invariant iff $%
\gamma =1/2$ we are interested in finding the relationship between $\Pi $
and $\rho $ (i.e. we are interesting in determining a new equation of state
relating $\Pi $ and $\rho $). For this purpose we try to integrate the
resulting ode under the restriction $\gamma =1/2,$ following the standard
Lie procedure but unfortunately we had not been able of obtaining any
explicit solution. Nevertheless we have obtained the invariant solution (a
particular solution) that induces the scaling symmetry finding in this way a
concrete power law solutions for the main quantities of the model. In this
way we arrive to the conclusion that $\Pi $ and $\rho $ has the same order
of magnitude and therefore we find that $\Pi =\varkappa \rho $ where $%
\varkappa $ is a negative numerical constant. These results are not new,
they have already been obtained by several author using different methods
and they will be commented in this section.

As we have been able to determine a concrete relationship between $\Pi $ and
$\rho $ (under the scale-invariant condition $\gamma =1/2$) in section 4 we
investigate if this condition implies $\gamma =1/2.$ For this purpose, under
this hypothesis, we obtain a second order ode that describes all the model
and when studying it with the Lie group method we find that such equation
only admits scaling symmetries iff $\gamma =1/2.$ We study the resulting ode
finding the same results than in the above section. Nevertheless, the
assumption $\Pi =\varkappa \rho ,$ allows us to obtain a complete solution
to the field equations, this possibility will be show in subsection 4.2.

In section 5 we again study the model as well as some of the odes that have
been arising in the paper through a pedestrian method, Dimensional Analysis.
In this section we shall show how this method works in order to obtain the
\textquotedblleft same\textquotedblright\ results but in a trivial way. We
end by summarizing some results.

\section{The model.}

For a flat Friedmann-Robertson-Walker (FRW) Universe with a line element
\begin{equation}
ds^{2}=c^{2}dt^{2}-f^{2}(t)\left( dx^{2}+dy^{2}+dz^{2}\right) ,
\end{equation}
filled with a bulk viscous cosmological fluid the energy-momentum tensor is
given by (see \cite{Maartens})
\begin{equation}
T_{i}^{k}=\left( \rho+p+\Pi\right) u_{i}u^{k}-\left( p+\Pi\right)
\delta_{i}^{k},  \label{1}
\end{equation}
where $\rho$ is the energy density, $p$ the thermodynamic pressure, $\Pi$
the bulk viscous pressure (stress) and $u_{i}$ the four velocity satisfying
the condition $u_{i}u^{i}=1$. The field equations yield:
\begin{align}
2H^{\prime}+3H^{2} & =-\kappa\left( p+\Pi\right) ,  \label{field1} \\
3H^{2} & =\kappa\rho,  \label{field2} \\
\rho^{\prime}+3\alpha\rho H & =-3H\Pi,  \label{field3} \\
\Pi^{\prime}+\frac{\Pi}{k_{\gamma}\rho^{\gamma-1}} & =-3\rho H-\frac{1}{2}%
\Pi\left( 3H-W\frac{\rho^{\prime}}{\rho}\right) ,  \label{field4}
\end{align}
where
\begin{equation}
H=\frac{f^{\prime}}{f},\qquad W=\frac{2\omega+1}{\omega+1}=1+\frac{\omega }{%
\alpha},\qquad\alpha=\left( \omega+1\right) ,\qquad\kappa=\frac{8\pi G}{c^{2}%
},
\end{equation}
and where we are assuming the following phenomenological (ad hoc) equation
of state (laws) for $p$, $\xi,$ $T$ and $\tau$ (see \cite{Maartens}):
\begin{equation}
p=\omega\rho,\quad\xi=k_{\gamma}\rho^{\gamma},\quad T=D_{\beta}\rho^{\beta
},\quad\tau=\xi\rho^{-1}=k_{\gamma}\rho^{\gamma-1},  \label{csi4}
\end{equation}
where $0\leq\omega\leq1$, and $k_{\gamma}\geq0$, $D_{\beta}>0$ are
dimensional constants, $\gamma\geq0$ and $\beta\geq0$ are numerical
constants. Eq. $\left( p=\omega\rho\right) $ is standard in cosmological
models whereas the equation for $\tau$ is a simple procedure to ensure that
the speed of viscous pulses does not exceed the speed of light. These
equations are introduced without sufficient thermodynamical motivation, but
in absence of better alternatives we shall follow the practice adopting them
in the hope that they will at least provide indication of the range of
possibilities. For the temperature law $T=D_{\beta}\rho^{\beta}$ which is
the simplest law guaranteeing positive heat capacity.

For a detailed deduction of this model see R. Maartens (\cite{Maartens})

\section{The General equation.}

Without any assumption the field eq. (\ref{field1}-\ref{field4}) may be
expressed by a single one

\begin{equation}
H^{\prime\prime}-K_{0}H^{-1}\left( H^{\prime}\right) ^{2}+K_{1}HH^{\prime
}+K_{2}H^{\prime}H^{2-2\gamma}+K_{3}H^{3}+K_{4}H^{4-2\gamma}=0,
\end{equation}
where
\begin{align*}
K_{0} & =W=\left( 1+\beta\right) ,\text{ \ \ \ \ }\beta=\frac{\omega }{%
\omega+1}, \\
K_{1} & =3\left( \alpha-\frac{\alpha W}{2}+\frac{1}{2}\right) =3, \\
K_{2} & =k_{\gamma}^{-1}\left( \frac{3}{\kappa}\right) ^{1-\gamma}=\frac{%
3^{1-\gamma}}{k_{\gamma}\kappa^{1-\gamma}},\qquad\left[ K_{2}\right]
=T^{1-2\gamma}, \\
K_{3} & =\frac{9}{2}\left( \frac{\alpha}{2}-1\right) =\frac{9}{4}\left(
\omega-1\right) , \\
K_{4} & =k_{\gamma}^{-1}\left( \frac{3}{\kappa}\right) ^{1-\gamma}\frac{%
3\alpha}{2}=3^{2-\gamma}\frac{\left( \omega+1\right) }{k_{\gamma
}\kappa^{1-\gamma}},\qquad\left[ K_{4}\right] =T^{1-2\gamma},
\end{align*}
since $\left[ k_{\gamma}\right] =L^{\gamma-1}M^{1-\gamma}T^{2\gamma -1},%
\left[ \kappa\right] =LM^{-1},$ hence: $\left[ K_{2}\right] =\frac {1}{%
T^{2\gamma-1}}.$

Taking into account the value of the constants $K_{i}$ this equations
yields:
\begin{equation}
H^{\prime\prime}-WH^{-1}\left( H^{\prime}\right)
^{2}+3HH^{\prime}+K_{2}H^{\prime}H^{2-2\gamma}+\frac{9}{4}\left(
\omega-1\right) H^{3}+K_{4}H^{4-2\gamma}=0,  \label{H-eq}
\end{equation}
and if we decide to make the following assumption $k_{\gamma}=\kappa=1$ then
eq. (\ref{H-eq}) yields:
\begin{equation}
H^{\prime\prime}-\left( \frac{2\omega+1}{\omega+1}\right) H^{-1}\left(
H^{\prime}\right) ^{2}+3HH^{\prime}+3^{1-\gamma}H^{\prime}H^{2-2\gamma}+%
\frac{9}{4}\left( \omega-1\right) H^{3}+\frac{3^{2-\gamma}}{2}\left(
\omega+1\right) H^{4-2\gamma}=0.  \label{DEF-H-eq}
\end{equation}

We go next to study this equation under the Lie Group technique (see for
example \cite{Ovsi}, \cite{Ibragimov} and \cite{Blumann}). For simplicity we
have rewrite it in the following form:
\begin{equation}
H^{\prime\prime}-AH^{-1}\left( H^{\prime}\right) ^{2}+3HH^{\prime
}+CH^{\prime}H^{2-2\gamma}+MH^{3}+EH^{4-2\gamma}=0.  \label{gc}
\end{equation}

The standard Lie procedure brings us to obtain the next system of pdes:
\begin{align}
\xi_{HH}H+A\xi_{H} & =0, \\
A\eta+6\xi_{H}H^{3}+\eta_{HH}H^{2}-2\xi_{tH}H^{2}-A\eta_{H}H+2C\xi
_{H}H^{4-2\gamma} & =0, \\
2\eta_{Ht}H-\xi_{tt}H+3\eta
H+3\xi_{t}H^{2}+3M\xi_{H}H^{4}+C\xi_{t}H^{3-2\gamma}+3E\xi_{H}H^{5-2\gamma}+%
\left( 2-2\gamma\right) C\eta H^{2-2\gamma}-2A\eta_{t} & =0, \\
3M\eta H+MH^{2}\left( 2\xi_{t}-\eta_{H}\right) +\left( 4-2\gamma\right)
E\eta H^{2-2\gamma}+C\eta_{t}H^{1-2\gamma}+\left( 2\xi_{t}-\eta_{H}\right)
EH^{3-2\gamma}+\eta_{tt}H^{-1}+3\eta_{t} & =0,
\end{align}
this system admits the following symmetry
\begin{equation}
\xi=1,\eta=0\Longrightarrow X_{1}=\partial_{t},
\end{equation}
since $X_{1}$ span an algebra $L_{1}$ the equation cannot be completely
integrated by the Lie group method.

But if we try to check if the system admits a scaling symmetry
\begin{equation}
\xi=t,\eta=-H,
\end{equation}
this is only possible iff $\gamma=1/2.$ Therefore
\begin{equation}
\xi=at+b,\eta=-aH,
\end{equation}
is a symmetry of the ode iff $\gamma=1/2,$ and where $a$ and $b$ are
numerical constants i.e. $a,b\in\mathbb{R}.$ Hence
\begin{equation}
X_{1}=\partial_{t},\qquad X_{2}=t\partial_{t}-H\partial_{H}\qquad\qquad\left[
X_{1},X_{2}\right] =X_{1},
\end{equation}
which span a solvable Lie algebra $L_{2}$ of the type $III.$

At the same result have arrived for example A.A.Coley et al (\cite{Coley}),
who study this model from a dynamical system approach. To apply this method
they rewrite the field equations in a dimensionless way in such a form that
this is only possible iff $\gamma=1/2$ and the viscous pressure and the
energy density has the same order of magnitude as we will see in the next
section. At similar conclusions R. A. Daishev and W. Zimdahl have arrived in
(\cite{Zindal}), where these authors study this model from the homothetic
(similarity) point of view i.e. they study when the field equations remain
self-similar. As we will see below we have obtained the same results but
using the Lie group method. Finally Belinch\'{o}n et al (\cite{Tony1}) have
obtained the same results using the renormalization group approach.

The canonical variables and the reduced ode that induces the symmetry $X_{1}$
are:
\begin{equation}
y(x)=\frac{1}{H^{\prime }},\qquad x=H,
\end{equation}%
\begin{equation}
y^{\prime }=\left( Mx^{3}+Ex^{4-2\gamma }\right) y^{3}+\left(
3x+Cx^{2-2\gamma }\right) y^{2}-A\frac{y}{x},
\end{equation}%
which is an Abel ode (see \cite{Odes}). Without any assumption it is very
difficult to find any explicit solution of this equation and therefore a
solution of eq. (\ref{gc})

\subsection{The case with $\protect\gamma=1/2,$ scale invariant solution.}

As we can see, the ode (\ref{gc}) admits a scaling symmetry iff $\gamma=1/2$
and in this case such ode is reduced to:
\begin{equation}
H^{\prime\prime}-WH^{-1}\left( H^{\prime}\right) ^{2}+\left( 3+\sqrt {3}%
\right) HH^{\prime}+\left( \frac{9}{4}\left( \omega-1\right) +\frac{3\sqrt{3}%
}{2}\left( \omega+1\right) \right) H^{3}=0,  \label{SI}
\end{equation}
or equivalently
\begin{equation}
H^{\prime\prime}-AH^{-1}\left( H^{\prime}\right) ^{2}+BHH^{\prime}+CH^{3}=0,
\label{SI reduced}
\end{equation}
where obviously it admits the symmetries
\begin{equation}
X_{1}=\partial_{t},\qquad X_{2}=t\partial_{t}-H\partial_{H},\qquad \qquad
\left[ X_{1},X_{2}\right] =X_{1},
\end{equation}
which form a $L_{2}$ algebra etc..

Symmetry $X_{1}$ brings us to the following ode through the reduction
(canonical variables)
\begin{equation}
y(x)=\frac{1}{H^{\prime}},\qquad x=H,
\end{equation}
\begin{equation}
y^{\prime}=\left( \frac{9}{4}\left( \omega-1\right) +\frac{3\sqrt{3}}{2}%
\left( \omega+1\right) \right) x^{3}y^{3}+\left( 3+\sqrt{3}\right) xy^{2}-W%
\frac{y}{x},
\end{equation}
\begin{equation}
y^{\prime}=Cx^{3}y^{3}+Bxy^{2}-A\frac{y}{x},  \label{abel1}
\end{equation}
which is an Abel ode. This ode admits the following symmetry
\begin{equation}
\tilde{X}=x\partial_{x}-2y\partial_{y},  \label{symabel1}
\end{equation}
which is a scaling symmetry and it induces the following change of
variables,
\begin{equation}
r=x^{2}y,\quad s(r)=\ln(x),\quad\Longrightarrow\quad x=e^{s(r)},\quad y=%
\frac{r}{e^{2s(r)}},
\end{equation}
which brings us to obtain the next ode in quadratures
\begin{equation}
s^{\prime}=\frac{1}{r\left( Cr^{2}+Br+2-A\right) },
\end{equation}
and which solution is:
\begin{equation}
s(r)=-\frac{\ln r}{A-2}+\frac{1}{2}\frac{\ln\left( Cr^{2}+Br+2-A\right) }{A-2%
}-\frac{B\arctan h\left( \frac{2Cr+B}{\sqrt{B^{2}+4C(A-2)}}\right) }{\left(
A-2\right) \sqrt{B^{2}+4C(A-2)}}+C_{1},
\end{equation}
and hence in the original variables $\left( x,y\right) $:
\begin{equation}
\ln x=-\frac{\ln\left( x^{2}y\right) }{A-2}+\frac{1}{2}\frac{\ln\left(
Cx^{4}y^{2}+Bx^{2}y+2-A\right) }{A-2}-\frac{B\arctan h\left( \frac {%
2Cx^{2}y+B}{\sqrt{B^{2}+4C(A-2)}}\right) }{\left( A-2\right) \sqrt {%
B^{2}+4C(A-2)}}+C_{1},
\end{equation}
therefore we have obtained the next ode in the $\left( H,t\right) $
variables:
\begin{equation}
\ln H=-\frac{\ln\left( \frac{H^{2}}{H^{\prime}}\right) }{A-2}+\frac{1}{2}%
\frac{\ln\left( C\left( \frac{H^{2}}{H^{\prime}}\right) ^{2}+B\left( \frac{%
H^{2}}{H^{\prime}}\right) +2-A\right) }{A-2}-\frac{B\arctan h\left( \frac{2C%
\frac{H^{2}}{H^{\prime}}+B}{\sqrt{B^{2}+4C(A-2)}}\right) }{\left( A-2\right)
\sqrt{B^{2}+4C(A-2)}}+C_{1},
\end{equation}
but we do not know how to obtain an ``explicit'' solution of this ode i.e. a
solution of the form $H=H(t)$. Possibly the most general solution to this
equation may result unphysical as we have pointed out in the case of a
perfect fluid (see the appendix of (\cite{Tony4})).

\subsubsection{Invariant solution}

In this case we can try to find a particular solution of eq. (\ref{SI})
through the invariant solution that induces the scaling symmetry $X_{2}=%
\left[ at,-aH\right] .$ In such case we find that
\begin{equation}
\frac{dt}{\xi}=\frac{dH}{\eta}\Longrightarrow H=\frac{a}{t},\qquad a\in%
\mathbb{R},
\end{equation}
which satisfies the eq. (\ref{SI}) iff $a=a(\omega),$ i.e.
\begin{equation}
a=\frac{B\pm\sqrt{B^{2}+4C(A-2)}}{2C}\Longleftrightarrow a=\frac{\left( 3+%
\sqrt{3}\right) \pm\sqrt{12-9\frac{\omega-1}{\omega+1}}}{\frac{9}{2}\left(
\omega-1\right) +3\sqrt{3}\left( \omega+1\right) },  \label{a}
\end{equation}

In this case, it is observed that
\begin{equation}
\rho=\frac{3}{\kappa}H^{2},\qquad\Pi=-\frac{2}{\kappa}H^{\prime}-\frac {%
3\alpha}{\kappa}H^{2}=\left( \frac{2}{3}-\alpha\right) \rho,
\end{equation}
therefore
\begin{equation}
\Pi\thickapprox\rho\Longrightarrow\Pi=\varkappa\rho,\qquad\varkappa=\left(
\frac{2}{3}-\alpha\right) \in\mathbb{R}^{-},
\end{equation}
i.e. we have found that the viscous pressure and the energy density have the
same order of magnitude and hence we can define a new equation of sate $%
\Pi=\varkappa\rho$ where for physical reasons the numerical constants $%
\varkappa$ must be negative. Note that if $\omega=1,$ then $\varkappa =-%
\frac{4}{3}.$ At the same result R. A. Daishev and W. Zimdahl have arrived (
\cite{Zindal}) through a very different way. Nevertheless in our solution it
is observed that $\varkappa$ can take other values.

Note that both quantities have the same dimensional equation i.e. $\left[ \Pi%
\right] =\left[ \rho\right] $ and that for this reason under a scaling
transformation the dimensionless quantity $\frac{\Pi}{\rho}$ must be remain
constant (see the pioneering work in this field of D. M. Eardly \cite%
{Eardley}, and the latter of K. Rosquits and R. Jantzen \cite{Jantzen}, and
J. Wainwright \cite{Wainwrit}). Under the action of a similarity, each
physical quantity $\phi$ transforms according to it dimension $q$ under
scale transformations i.e. changes of the unit of length. Thus if unit of
length $L$ transforms as $L\longrightarrow\lambda L$ then $%
\phi\longrightarrow\lambda ^{q}\phi.$ This means that dimensionless
quantities are invariant under a similarity transformation. Dimensionless
quantities are therefore spacetimes constants. This implies that two
quantities with the same dimensions, for example $\Pi$ and $\rho$ or $p$ and
$\rho$ are related through equations of state of the form $\Pi=\varkappa\rho$
or $p=\omega\rho$ since the ratios $\frac{\Pi}{\rho}$ or $\frac{p}{\rho}$
must be constants. Furthermore, as J. Wainwright have pointed out,
spacetimes admitting transitively self-similarity groups correspond exactly
to the exact power law solutions as we have found.

In the same way we can try to find a particular solution of eq.(\ref{abel1})
that induces the symmetry $\tilde{X}=x\partial_{x}-2y\partial_{y}$.
Therefore we find that
\begin{equation}
\frac{dx}{x}=-\frac{dy}{2y}\Longrightarrow y=\frac{\tilde{a}}{x^{2}},
\end{equation}
is a solution of eq. (\ref{abel1}) iff
\begin{equation}
\tilde{a}=\frac{-B\pm\sqrt{B^{2}+4C(A-2)}}{2C},
\end{equation}
note that $\tilde{a}=-a.$ Now tanking into account the change of variables $%
\left( y(x)=\frac{1}{H^{\prime}},x=H\right) $ it is found that
\begin{equation}
\frac{1}{H^{\prime}}=\frac{\tilde{a}}{H^{2}}\Longrightarrow H=\frac{a}{t},
\end{equation}
where $a$ is given by eq. (\ref{a}).

\subsubsection{The case with $\protect\gamma=1/2,$ scale invariant solution
and $\protect\omega=1,$ stiff matter.}

We shall study the equation
\begin{equation}
H^{\prime\prime}-\frac{3}{2}H^{-1}\left( H^{\prime}\right) ^{2}+\left( 3+%
\sqrt{3}\right) HH^{\prime}+3\sqrt{3}H^{3}=0,  \label{scaling}
\end{equation}
which is a special case of eq. (\ref{SI}). (canonical variables)
\begin{equation}
y(x)=\frac{1}{H^{\prime}},\qquad x=H,
\end{equation}
\begin{equation}
y^{\prime}=3\sqrt{3}x^{3}y^{3}+\left( 3+\sqrt{3}\right) xy^{2}-\frac{3}{2}%
\frac{y}{x},
\end{equation}
which is an Abel ode. The following change of variables brings us to obtain
the next new ode
\begin{equation}
s(r)=-2\ln x,r=yx^{2}\quad\Longrightarrow\quad x=e^{-1/2s(r)},y=\frac {r}{%
e^{-1/s(r)}},
\end{equation}
\begin{equation}
s^{\prime}=-\frac{4}{r\left( 1+3\sqrt{3}r^{2}+2\left( 3+\sqrt{3}\right)
r\right) },
\end{equation}
and which solution is:
\begin{equation}
s=\frac{2\sqrt{3}\ln\left( 6r+\sqrt{3}-1\right) }{\sqrt{3}-1}-\frac {8\sqrt{3%
}\ln r}{\left( \sqrt{3}-1\right) \left( \sqrt{3}+3\right) }-\frac{2\sqrt{3}%
\ln\left( 6r+\sqrt{3}+3\right) }{\sqrt{3}+3},
\end{equation}
\begin{equation}
-2\ln x=\frac{2\sqrt{3}\ln\left( 6yx^{2}+\sqrt{3}-1\right) }{\sqrt{3}-1}-%
\frac{8\sqrt{3}\ln\left( yx^{2}\right) }{\left( \sqrt{3}-1\right) \left(
\sqrt{3}+3\right) }-\frac{2\sqrt{3}\ln\left( 6yx^{2}+\sqrt {3}+3\right) }{%
\sqrt{3}+3},
\end{equation}
and hence
\begin{equation}
-2\ln H=\frac{2\sqrt{3}\ln\left( 6\frac{H^{2}}{H^{\prime}}+\sqrt{3}-1\right)
}{\sqrt{3}-1}-\frac{8\sqrt{3}\ln\left( \frac{H^{2}}{H^{\prime}}\right) }{%
\left( \sqrt{3}-1\right) \left( \sqrt{3}+3\right) }-\frac{2\sqrt{3}\ln\left(
6\frac{H^{2}}{H^{\prime}}+\sqrt{3}+3\right) }{\sqrt{3}+3},
\end{equation}

\textbf{The invariant solution} that we can find in this case is:
\begin{equation}
H=\frac{a}{t},\qquad a\in\mathbb{R},
\end{equation}
which satisfies the eq. (\ref{scaling}) iff
\begin{equation}
a=\frac{\left( 3+\sqrt{3}\right) \pm\sqrt{12}}{6\sqrt{3}}=\left\{
\begin{array}{c}
\frac{1}{2}+\frac{\sqrt{3}}{6}=0.7886751351, \\
-\frac{1}{6}+\frac{\sqrt{3}}{6}=0.122008468.%
\end{array}
\right.
\end{equation}

\section{The case $\Pi=\varkappa\protect\rho.$}

Since under a scale transformation we have found that the viscous parameter
must be $\gamma=1/2,$ and that in such case $\Pi=\varkappa\rho,$ now we are
interested in studying the inverse way i.e. if under the hypothesis $%
\Pi=\varkappa\rho,$ the resulting differential equation remains scale
invariant. For this purpose we rewrite the field eq. (\ref{field1}-\ref%
{field4}) under the assumption $\Pi=\varkappa\rho.$ In this way the field
equations may be expressed by the following ode:

\begin{equation}
\rho^{\prime\prime}=\frac{\rho^{\prime2}}{\rho}-A\beta\rho^{\beta}\rho
^{\prime}+B\rho^{2},  \label{neweq1}
\end{equation}
with $\left( 1-\gamma\right) =\beta$ and
\begin{align*}
A & =D^{-1}k_{\gamma}^{-1}=\frac{2\left( \omega+1\right) }{k_{\gamma}},\qquad%
\left[ A\right] =\left[ k_{\gamma}\right] ^{-1}=L^{1-\gamma
}M^{\gamma-1}T^{1-2\gamma}, \\
B & =\frac{\kappa\varpi}{2\delta D}=\frac{\kappa\left( \omega+1+\varkappa
\right) \left( \omega+1\right) \left( 6+3\varkappa\right) }{2\varkappa }%
,\qquad\left[ B\right] =\left[ \kappa\right] =LM^{-1},
\end{align*}
\begin{equation*}
\varpi=(\alpha+\varkappa),\qquad\alpha=\left( \omega+1\right) ,\text{\ \ \ }%
\delta=\frac{2\varkappa}{6+3\varkappa},\text{ \qquad\ }D=\left( 1-\frac{W}{2}%
\right) =\frac{1}{2\left( \omega+1\right) }.
\end{equation*}

The Lie analysis of equation (\ref{neweq1}) brings us to obtain \ the
following system of pdes
\begin{align}
\xi _{\rho \rho }+\rho ^{-1}\xi _{\rho }& =0,  \label{cons1} \\
\eta \rho ^{-2}-\eta _{\rho }\rho ^{-1}+\left( \eta _{\rho \rho }-2\xi
_{t\rho }\right) +2\xi _{\rho }A\beta \rho ^{\beta }& =0,  \label{cons2} \\
\left( 2\eta _{t\rho }-\xi _{tt}\right) +\xi _{t}A\beta \rho ^{\beta }-3\xi
_{\rho }B\rho ^{2}-\eta _{t}2\rho ^{-1}+\eta A\beta ^{2}\rho ^{\beta -1}& =0,
\label{cons3} \\
-\eta 2B\rho +\eta _{tt}+\eta _{\rho }B\rho ^{2}-2\xi _{t}B\rho ^{2}+\eta
_{t}A\beta \rho ^{\beta }& =0,  \label{cons4}
\end{align}%
we solve eqs. (\ref{cons1}-\ref{cons4}), finding that this system only
admits the symmetry
\begin{equation}
X_{1}=\partial _{t}.
\end{equation}

Now, if we try to check if the system admits a scaling symmetry
\begin{equation}
X_{2}=a\beta t\partial _{t}+a\rho \partial _{\rho },\qquad a\in \mathbb{R},
\end{equation}%
we see with the help of eq. (\ref{cons4}) that
\begin{equation}
-2Aa\rho ^{2}+Aa\rho ^{2}+2a\beta A\rho ^{2}=0,
\end{equation}%
finding in this way that
\begin{equation}
-1+2\beta =0\Longleftrightarrow \beta =\frac{1}{2},  \label{beta}
\end{equation}%
where $\beta =\left( 1-\gamma \right) ,$ that is to say $\gamma =\frac{1}{2}.
$
\begin{equation}
\xi (\rho ,t)=-\frac{a}{2}t+b,\text{ \ \ \ \ \ \ \ \ \ \ }\eta (\rho
,t)=a\rho ,
\end{equation}

Therefore as we can see the ode only admits a single symmetry $%
X_{1}=\partial_{t},$ but if we impose that the ode admits a scaling
symmetry, we have seen that this is only possible if $\gamma=\frac{1}{2}.$
Therefore iff $\gamma=\frac{1}{2}$ the ode admits two symmetries:
\begin{equation}
X_{1}=\partial_{t},\text{ \ \ \ \ \ \ \ }X_{2}=t\partial_{t}-2\rho
\partial_{\rho},\qquad\qquad\left[ X_{1},X_{2}\right] =X_{1},
\end{equation}
where $X_{2}$ is the generator of the scaling group. Hence we can conclude
that the assumption $\Pi=\varkappa\rho$ does not imply that the resulting
field equations must be scale invariant, this is only possible if $\gamma=%
\frac{1}{2}.$

The symmetry $X_{1}$ brings us to obtain through the canonical variables the
following Abel ode:
\begin{equation}
y^{\prime}=-Bx^{2}y^{3}+A\beta x^{\beta}y^{2}-\frac{y}{x},  \label{abel}
\end{equation}
where
\begin{equation}
x=\rho,\qquad y=\frac{1}{\rho^{\prime}}.
\end{equation}

\subsection{Equation (\protect\ref{neweq1}) with $\protect\gamma=1/2.$}

The equation (\ref{neweq1}) with $\gamma=1/2$ yields
\begin{equation}
\rho^{\prime\prime}=\frac{\rho^{\prime2}}{\rho}-\frac{A}{2}\sqrt{\rho}%
\rho^{\prime}+B\rho^{2},  \label{newecu}
\end{equation}
where $\left[ A^{2}\right] =\left[ B\right] .$

The symmetry $X_{1}$ brings us to obtain through the canonical variables the
following Abel ode:
\begin{equation}
y^{\prime}=-Bx^{2}y^{3}+\frac{A}{2}\sqrt{x}y^{2}-\frac{y}{x},
\label{newAbel}
\end{equation}
where
\begin{equation}
x=\rho,\qquad y=\frac{1}{\rho^{\prime}},  \label{ainoa}
\end{equation}

We would like to point out that eq. (\ref{newAbel}) admits the following
scaling symmetry
\begin{equation}
\tilde{X}=x\partial_{x}-\frac{3}{2}y\partial_{y},  \label{symnewAbel}
\end{equation}
which induces the following change of variables
\begin{equation}
r=yx^{3/2},\text{ \ \ }s(r)=\ln x\text{ \qquad}\Longrightarrow\qquad y=\frac{%
r}{e^{3s(r)/2}},\text{ \ }x=e^{s(r)},
\end{equation}
in such a way that eq. (\ref{newAbel}) yields
\begin{equation}
s^{\prime}=\frac{2}{r\left( 1+rA-2r^{2}B\right) },  \label{quadrature}
\end{equation}
where the solution of eq. (\ref{quadrature}) is:
\begin{equation}
s(r)=2\ln r-\ln\left( 1+rA-2r^{2}B\right) +\frac{2\arctan h\left( \frac{A-4rB%
}{\sqrt{A^{2}+8B}}\right) }{\sqrt{A^{2}+8B}}+C_{1},
\end{equation}
hence
\begin{equation}
\ln x=2\ln\left( yx^{3/2}\right) -\ln\left( 1+\left( yx^{3/2}\right)
A-2\left( y^{2}x^{3}\right) B\right) +\frac{2\arctan h\left( \frac{A-4\left(
yx^{3/2}\right) B}{\sqrt{A^{2}+8B}}\right) }{\sqrt {A^{2}+8B}}+C_{1},
\end{equation}
and tanking into account the change of variables (\ref{ainoa}) yields
\begin{equation}
\ln\rho=2\ln\left( \frac{\rho^{3/2}}{\rho^{\prime}}\right) -\ln\left(
1+\left( \frac{\rho^{3/2}}{\rho^{\prime}}\right) A-2\left( \frac{\rho^{3}}{%
\rho^{\prime2}}\right) B\right) +\frac{2\arctan h\left( \frac{A-4\left(
\frac{\rho^{3/2}}{\rho^{\prime}}\right) B}{\sqrt{A^{2}+8B}}\right) }{\sqrt{%
A^{2}+8B}}+C_{1},
\end{equation}
which is a quadrature, but unfortunately we do not know how to obtain an
``explicit'' solution for this ode as in the above case.

\subsubsection{Invariant solution}

The invariant solution is obtained for $a\neq0$ For this value of $a$ eq. (%
\ref{newecu}) admits a single symmetry
\begin{equation}
\xi\left( t,\rho\right) =at\partial_{t},\qquad\eta\left( x,y\right)
=-2a\rho\partial_{\rho},
\end{equation}
the knowledge of one symmetry $X$ might suggest the form of a particular
solution as an invariant of the operator $X$ i.e. the solution of
\begin{equation}
\frac{dt}{\xi\left( t,\rho\right) }=\frac{d\rho}{\eta\left( t,\rho\right) },
\label{ecu7}
\end{equation}
this particular solution is known as an invariant solution (generalization
of similarity solution). In this case
\begin{equation}
\rho=\rho_{0}t^{-2},\text{ \ \ \ \ \ \ \ \ / \ \ \ }\rho_{0}=\frac{1}{2}%
\frac{4B+A^{2}\pm A\sqrt{\left( 8B+A^{2}\right) }}{B^{2}},
\end{equation}
where
\begin{equation}
A=2\left( \omega+1\right) ,\qquad B=\frac{\left( \omega+1+\varkappa\right)
\left( \omega+1\right) \left( 6+3\varkappa\right) }{2\varkappa},
\end{equation}
with $k_{\gamma}=\kappa=1$ and making $\omega=1$%
\begin{equation}
A=4,\qquad B=3\frac{\left( 2+\varkappa\right) ^{2}}{\varkappa},
\end{equation}

A particular solution of eq. (\ref{newAbel}) may be found by taking into
account the symmetry $\tilde{X}=x\partial _{x}-\frac{3}{2}y\partial _{y}.$
In this case
\begin{equation}
3\frac{dx}{x}=-2\frac{dy}{y}\Longrightarrow y=\frac{a}{x^{3/2}},
\end{equation}%
and taking into account the change of variables $\left( x=\rho ,y=\frac{1}{%
\rho ^{\prime }}\right) $ it is founded the already known solution $\rho
=\rho _{0}t^{-2}.$

\subsection{The General solution.}

In this case, the assumption $\Pi =\varkappa \rho $ allows us to obtain a
complete solution to the field equations (\ref{field1}-\ref{field4}).

If we take into account eq. (\ref{field4}) with the assumption $\Pi
=\varkappa \rho ,$ it yields:%
\begin{equation}
\varkappa \rho ^{\prime }+\varkappa k_{\gamma }^{-1}\rho ^{\gamma -2}=-\frac{%
1}{\left( \alpha +\varkappa \right) }\rho ^{\prime }+\frac{\varkappa }{2}%
\frac{1}{\left( \alpha +\varkappa \right) }\rho ^{\prime }+\frac{\varkappa W%
}{2}\rho ^{\prime },  \label{melissa}
\end{equation}%
where $H$ has been obtained from eq. (\ref{field3}) and follows the
relationship
\begin{equation}
H=-\frac{1}{3\left( \alpha +\varkappa \right) }\frac{\rho ^{\prime }}{\rho }.
\end{equation}

Simplifying eq. (\ref{melissa}) it yields%
\begin{equation}
\left( \varkappa +\frac{1}{\alpha +\varkappa }-\frac{\varkappa }{2\left(
\alpha +\varkappa \right) }-\frac{\varkappa W}{2}\right) \rho ^{\prime
}=-\varkappa k_{\gamma }^{-1}\rho ^{\gamma -2},  \label{china1}
\end{equation}%
which trivial solution is:%
\begin{equation}
\rho =\rho _{0}t^{-\frac{1}{1-\gamma }},  \label{sol-mel}
\end{equation}%
where
\begin{equation}
\rho _{0}=\frac{-\varkappa k_{\gamma }^{-1}}{\left( \varkappa +\frac{1}{%
\alpha +\varkappa }-\frac{\varkappa }{2\left( \alpha +\varkappa \right) }-%
\frac{\varkappa W}{2}\right) }.
\end{equation}

In this way we have obtained a complete solution for the field equations (%
\ref{field1}-\ref{field4}) and valid for all value of $\gamma .$ It is
obvious that when $\gamma =1/2$ then we recover our previous solution $\rho
=\rho _{0}t^{-2}.$

\section{A Pedestrian Method.}

In this section we would like to show how dimensional Analysis works in
order to obtain the same results but in a trivial way (see for example \cite%
{Blarenblat}, \cite{Kurt}, \cite{Palacios}, and \cite{Tony2}). In the first
place we will show how by writing the field equations in a dimensionless way
we can determine the exact value of the parameter $\gamma$ which remains the
equations scale invariant. In second place we would like to show how to
solve some of the different odes that have arisen in this paper.

Writing the field equations (\ref{field1}-\ref{field4}) in a dimensionless
way and taken into account the following equations of state (\ref{csi4}) it
is obtained the following $\pi-monomia$ (see \cite{Castañs}):
\begin{align}
\pi_{1} & =\frac{Gpt^{2}}{c^{2}},\text{ \ \ \ \ \ \ \ }\pi_{2}=\frac{G\Pi
t^{2}}{c^{2}},\text{ \ \ \ \ \ \ \ \ }\pi_{3}=\frac{G\rho t^{2}}{c^{2}},%
\text{ \ \ }\pi_{4}=\frac{\Pi}{p}\text{ \ \ \ \ \ \ }\pi_{5}=\frac{\xi}{\Pi t%
},  \label{pi2} \\
\pi_{6} & =\frac{\tau}{t}=\tau H\text{, \ \ \ \ }\pi_{7}=\frac{\xi}{%
k_{\gamma}\rho^{\gamma}},\text{ \ \ \ \ \ \ \ \ }\pi_{8}=\frac{\xi}{\tau\rho
},\text{ \ \ \ \ \ \ \ }\pi_{9}=\frac{T}{D_{\beta}\rho^{\beta}},\text{\ \ \ }%
\pi_{10}=\frac{\rho}{p},  \label{pi4}
\end{align}
it is observed that from $\pi_{7}=\frac{\xi}{k_{\gamma}\rho^{\gamma}}$ and $%
\pi_{8}=\frac{\xi}{\tau\rho}$ we obtain
\begin{equation}
\widetilde{\pi}_{8}=\frac{k_{\gamma}\rho^{\gamma-1}}{t},\text{ }
\end{equation}
and from $\pi_{3}$ and $\widetilde{\pi}_{8}$%
\begin{equation}
\frac{G\rho t^{2}}{c^{2}}=\frac{k_{\gamma}\rho^{\gamma-1}}{t}\text{ \ \ \ }%
\Longrightarrow\qquad\rho=\left( \frac{t}{k_{\gamma}}\right) ^{1/\left(
\gamma-1\right) },
\end{equation}
therefore
\begin{equation}
\frac{c^{2}}{Gt}=\left( \frac{t}{k_{\gamma}}\right) ^{1/\left(
\gamma-1\right) }\Longrightarrow\qquad\frac{k_{\gamma}^{b}c^{2}}{G}=t^{b+2},
\end{equation}
where $b=1/(\gamma-1)$, it is observed that the only case and only for this,
$\gamma=1/2,$ we obtain the relationship $k_{\gamma}^{2}=c^{2}/G.$ If $%
\gamma\neq1/2$ the ``constants'' $G$ or $c$ must vary or we need to impose
the condition $G/c^{2}=const.$ (if both constants vary) if we want our
equations to remain scale invariant as we have showed in an earlier paper
(see \cite{Tony3}) where we studied a viscous model with $G$ time-varying.
In such work we arrived to the conclusion that if $\gamma=1/2, $ $G$ must be
constant in spite of considering it as a function that vary on time $t,$
since we were only interested in the self-similar solution of that model.

Now we go next to solve some of the differential equations that have arisen
in this paper through the Dimensional technique.

We begin studying eq. (\ref{SI reduced}) i.e.
\begin{equation}
H^{\prime\prime}-AH^{-1}\left( H^{\prime}\right) ^{2}+BHH^{\prime}+CH^{3}=0
\end{equation}
which verifies the principle of dimensional homogeneity taking into account
the dimensional base $B=\left\{ T\right\} .$ In this case we trivially
arrive to the solution $H\propto t^{-1}$ since $\left[ H\right] =T^{-1}.$
Note that D.A. (Pi theorem) does not understand numerical constants only of
orders of magnitude.

In second place we study eq. (\ref{abel1})
\begin{equation}
y^{\prime}=Cx^{3}y^{3}+Bxy^{2}-A\frac{y}{x},
\end{equation}
with respect to the dimensional base $B=\left\{ T\right\} .$ This ode
verifies the principle of dimensional homogeneity with respect to this
dimensional base. Note that $\left[ y\right] =\left[ \frac{1}{H^{\prime}}%
\right] =T^{2},$ and $\left[ x\right] =\left[ H\right] =T^{-1}$ hence $\left[
y^{\prime}\right] =T^{3}.$ Therefore rewriting the equation in a
dimensionless way \ we find that $y\propto x^{-2}$

But if we study this equation with respect to the dimensional base $%
B=\left\{ X,Y\right\} ,$ we need to introduce new dimensional constants that
make that the equation verifies the principle of dimensional homogeneity
\begin{equation}  \label{tetas}
y^{\prime}=\alpha Cx^{3}y^{3}+\beta Bxy^{2}-A\frac{y}{x}
\end{equation}
where $\left[ \alpha^{1/2}\right] =\left[ \beta\right] =X^{-2}Y^{-1},$ hence
\begin{equation}
\begin{array}{r|rrr}
& y & \beta & x \\ \hline
X & 0 & -2 & 1 \\
Y & 1 & -1 & 0%
\end{array}
\Longrightarrow y\propto\frac{\beta}{x^{2}},
\end{equation}

As we can see we have obtained the same solution than in the case of the
invariant solution. This is because the invariant solution that induces a
scaling symmetry is the same as the obtained one through the Pi theorem.

We would like to emphasize that D.A. brings us to obtain change of variables
(c.v.) (see \cite{Tony odes} for more details) which allows us to obtain
odes simplest than the original one. in this case, it is observed that $%
\left[ \beta\right] =X^{-2}Y^{-1}$ in such a way that we have the c.v.
\begin{equation}
\left( t=x,\qquad u(t)=\beta x^{2}y\right) \Longrightarrow\left( x=t,\qquad
y=\frac{u}{\beta t^{2}}\right) ,
\end{equation}
therefore eq. (\ref{tetas}) yields:
\begin{equation}
tu^{\prime}=u\left( u^{2}+u+1\right) ,
\end{equation}
and hence
\begin{equation}
\ln t+\frac{1}{2}\ln\left( u^{2}+u+1\right) +\frac{\sqrt{3}}{3}\arctan\left(
\left( \frac{3}{2}u+\frac{1}{3}\right) \sqrt{3}\right) -\ln u+C_{1}=0,
\end{equation}
in the original variables it yields
\begin{equation}
\ln x+\frac{1}{2}\ln\left( \left( ax^{2}y\right) ^{2}+ax^{2}y+1\right) +%
\frac{\sqrt{3}}{3}\arctan\left( \left( \frac{3}{2}\left( ax^{2}y\right) +%
\frac{1}{3}\right) \sqrt{3}\right) -\ln\left( ax^{2}y\right) +C_{1}=0.
\end{equation}

In the same way we can study eq. (\ref{newecu}) $.$
\begin{equation}
\rho^{\prime\prime}=\frac{\rho^{\prime2}}{\rho}-\frac{A}{2}\sqrt{\rho}%
\rho^{\prime}+B\rho^{2},
\end{equation}
where $\left[ A^{2}\right] =\left[ B\right] =LM^{-1}$, with respect to the
dimensional base $B=\left\{ L,M,T\right\} .$ Therefore it is found
\begin{equation*}
\begin{array}{r|rrr}
& \rho & B & t \\ \hline
L & -1 & 1 & 0 \\
M & 1 & -1 & 0 \\
T & -2 & 0 & 1%
\end{array}
\Longrightarrow\rho\propto\frac{1}{Bt^{2}}.
\end{equation*}

To end we study eq. (\ref{newAbel})
\begin{equation}
y^{\prime}=-Bx^{2}y^{3}+\frac{A}{2}\sqrt{x}y^{2}-\frac{y}{x},
\end{equation}
where $\left[ A^{2}\right] =\left[ B\right] =X^{-3}Y^{-2}$, with respect to
the dimensional base $B=\left\{ X,Y\right\} .$ Therefore it is found
\begin{equation}
\begin{array}{r|rrr}
& y & B & x \\ \hline
X & 0 & -2 & 1 \\
Y & 1 & -1 & 0%
\end{array}
\Longrightarrow y\propto\sqrt{\frac{1}{Bx^{3}}},
\end{equation}
as already we know.

\section{Conclusions}

In this paper we have studied the possible symmetries that admits a flat FRW
model filled with a bulk viscous fluid. Using the Lie group method we have
tried to find an adequate equation of state for the viscous parameter as
well as for the viscous pressure. Therefore we conclude that the field
equations remain scale invariant iff $\gamma =1/2$ and that for this
parameter it is found that $\Pi =\varkappa \rho .$ But that the hypothesis $%
\Pi =\varkappa \rho $ does not imply that the field equations remain scale
invariant, this only occurs if $\gamma =1/2.$Furthermore, the assumption $%
\Pi =\varkappa \rho $ brings us to obtain a complete solution to the field
equations valid for all $\gamma .$

\begin{acknowledgement}
I would wish to express my gratitude to M. Szydlowsky, R. Jantzen and J.
Wainwright for showing me their works and to Javier Aceves for his
translation into English of this paper.
\end{acknowledgement}


\begin{thebibliography}{99}
\bibitem{Collins} {\footnotesize C.B. Collins, \textit{Gen. Rel. Grav}.
\textbf{8} (1977) 717. }

{\footnotesize C.B. Collins, \textit{J. Math. Phys}. \textbf{18} (1977)
1374. }

\bibitem{Polacos} {\footnotesize M. Szydlowsky and M. Heller. \textit{Acta
Physica Polonica B}\textbf{14} (1983) 571. }

{\footnotesize M. Biesiada, M. Szydlowsky and J. Szczesny. \textit{Acta
Cosmologica }\textbf{16} (1989) 115. }

\bibitem{Maartens} {\footnotesize R. Maartens, \textit{Class. Quantum Grav.,
\textbf{12}} (1995) 1455. }

{\footnotesize R. Maartens, astro-ph/9609119. }

\bibitem{Ovsi} {\footnotesize L.V. Ovsiannkov. ``\textit{Group Analysis of
ODEs}''. Acd. Press. (1982). }

\bibitem{Ibragimov} {\footnotesize N.H. Ibragimov. \textquotedblleft\textit{%
Introduction to Modern Group Analysis}\textquotedblright. Ufa 2000. }

{\footnotesize N.H. Ibragimov. \textquotedblleft\textit{Elementary Lie Group
Analysis and ODEs}\textquotedblright. John Wiey \& Sons (1999). }

\bibitem{Blumann} {\footnotesize G.W. Blumann and S.C. Anco. ``\textit{%
Symmetry and Integration Methods for Differential equations}''.
Springer-Verlang (2002). }

\bibitem{Coley} {\footnotesize A.A. Coley, R.J. van den Hoogen and R.
Maartens. \textit{Phys. Rev D} \textbf{54} (1996) 1393. }

\bibitem{Zindal} {\footnotesize R. A. Daishev and W. Zimdahl. \textit{Class.
Quantum Grav., }\textbf{20} (2003) 5017. }

\bibitem{Tony1} {\footnotesize J.A. Belinch\'{o}n, T. Harko and M.K. Mak,
\textit{Class. Quantum Grav., \textbf{19}} (2002) 3003. }

\bibitem{Odes} {\footnotesize A.D. Polyanin \& V.F. Zaitsev. ``\textit{%
Handbook of Exact Solutions for ODEs}''. CRC Press (1995). }

\bibitem{Tony4} {\footnotesize J.A. Belinch\'{o}n, submitted to Jour.
Franklin Institute. qr-qc/0404028.}

\bibitem{Eardley} {\footnotesize D. M. Eardly. Commun .Math. Phys. 37 (1974)
287. }

\bibitem{Jantzen} {\footnotesize K. Rosquits and R. Jantzen, \textit{Class.
Quantum Grav., \textbf{2}} (1985) L129. }

{\footnotesize K. Rosquits and R. Jantzen, ``\textit{Transitively
Self-Similarity Space-Times}''. Proc. Marcel Grossmann Meeting on General
Relativity. Ed. Ruffini. Elsevier S.P. (1986). pg 1033. }

\bibitem{Wainwrit} {\footnotesize J. Wainwright, ``Self-Similar Solutions of
Einstein%
\'{}%
s Equations''. Published in Galaxies, Axisymmetric Systems \& Relativity. ed
M.A.H MacCallum CUP (1985). }

{\footnotesize J. Wainwright, \textit{Gen. Rel. Grav}. \textbf{16} (1984)
657.}

\bibitem{Blarenblat} {\footnotesize Barenblat. ``\textit{Scaling,
self-similarity and intermediate asymptotics}''. Cambridge texts in applied
mathematics N 14 1996 CUP. }

\bibitem{Kurt} {\footnotesize K. Kurth. ``\textit{Dimensional Analysis and
Group Theory in Astrophysics}''. Pergamon Press Oxford UK (1972). }

\bibitem{Palacios} {\footnotesize J. Palacios. ``\textit{Dimensional
Analysis''}. Macmillan 1964 London. }

\bibitem{Tony2} {\footnotesize J.A. Belinch\'{o}n, physics/9811016 }

\bibitem{Castañs} {\footnotesize M. Casta\~{n}s, ``\textit{Sobre el primer
postulado de Palacios}.'' JTGAD XXXI 1995.Preprint Group of D.A. Dept. of
Physic ETS Architecture UPM }

\bibitem{Tony3} {\footnotesize J.A. Belinch\'{o}n, T. Harko and M.K. Mak,
\textit{Gra. \& Cos}. \textbf{8} (2002) 319. }

\bibitem{Tony odes} {\footnotesize J.A. Belinch\'{o}n, physics/0502154}
\end{thebibliography}
\end{document}